\documentclass[a4paper,fleqn,usenatbib]{mnras}

\usepackage[T1]{fontenc}
\usepackage{ae,aecompl}

\usepackage{graphicx}	
\usepackage{amsmath}	
\usepackage{amssymb}	

\title[BPASS: Binary Black-Hole Mergers]{BPASS predictions for Binary Black-Hole Mergers}

\author[J.J. Eldridge \& E.R. Stanway]{
J.J. Eldridge,$^{1}$\thanks{E-mail: j.eldridge@auckland.ac.nz} and
E. R. Stanway,$^{2}$\thanks{E-mail: e.r.stanway@warwick.ac.uk}\\
$^{1}$Department of Physics, University of Auckland, Private Bag 92019, Auckland, New Zealand.\\
$^{2}$Department of Physics, University of Warwick, Gibbet Hill Road, Coventry, CV4 7AL, UK.\\
}

\date{Accepted XXX. Received YYY; in original form ZZZ}

\pubyear{2016}

\begin{document}
\label{firstpage}
\pagerange{\pageref{firstpage}--\pageref{lastpage}}
\maketitle

\begin{abstract}
Using the Binary Population and Spectral Synthesis code BPASS, we have
calculated the rates, timescales and mass distributions for binary
black hole mergers as a function of metallicity. We consider these in
the context of the recently reported 1st LIGO event detection. We find
that the event has a very low probability of arising from a stellar
population with initial metallicity mass fraction above $Z=0.010$
($Z\ga0.5\,Z_\odot$). Binary black hole merger events with the
reported masses are most likely in populations below 0.008
($Z\la0.4\,Z_\odot$).  Events of this kind can occur at all stellar
population ages from ~3 Myr up to the age of the universe, but
constitute only 0.1 to 0.4 per cent of binary BH mergers between
metallicities of $Z=0.001$ to 0.008. However at metallicity
$Z=10^{-4}$, 26 per cent of binary BH mergers would be expected to
have the reported masses. At this metallicity the progenitor merger
times can be close to $\approx 10$Gyr and rotationally-mixed stars
evolving through quasi-homogeneous evolution, due to mass transfer in
a binary, dominate the rate.  The masses inferred for the black holes
in the binary progenitor of GW\,150914 are amongst the most massive
expected at anything but the lowest metallicities in our models.  We
discuss the implications of our analysis for the electromagnetic
follow-up of future LIGO event detections.
\end{abstract}




\section{Introduction}

The recent detection of a gravitational wave transient from the
inspiral of a binary black-hole system \citep{discovery} opens a new
era in observations of the Universe.  While the existence of
gravitational waves had been inferred from observations of binary
pulsar systems \citep{1975ApJ...195L..51H}, for the first time,
ground-based, laser interferometeric experiments have made it possible
to observe a binary black hole system (BH-BH) merger and infer its
parameters independently of electromagnetic observations.

The detection of a gravitational wave transient, GW\,150914, by the
Laser Interferometer Gravitational-Wave Observatory
\citep[LIGO,][]{2015CQGra..32g4001L} was reported on 11th Feb
2016. Detected on 14th September 2015, during the first advanced-LIGO
operational run, the transient's characteristics are consistent with
the inspiral, merger and ring-down of a binary system comprising two
black holes, with estimated masses of $36^{+5}_{-4}$ and
$29^{+4}_{-4}$\,M$_{\odot}$ \citep{properties}. This event constitutes
the first detection of gravitational wave emission, and is notable in
that it represents a binary black hole (BH-BH) merger, rather than the
neutron star-neutron star (NS-NS) mergers expected to be more common
(if less luminous) sources at LIGO frequencies and sensitivities
\citep[e.g.][]{2010CQGra..27q3001A,Berry15}, although it was predicted
by some that BH-BH mergers would be more likely
\citep{Belczynski10,Dominik2015}. The event was not securely localised
in electromagnetic follow-up \citep[although
  see][]{2016arXiv160203920C}, and hence its host galaxy, and the
stellar population that generated it, remains unidentified.

Nearly all the black holes in the Universe are thought to be the end
result of stellar evolution. The only other source may be primordial
black holes formed during the Big Bang; these would only occur if
density perturbations are great enough that gravitational collapse
would occur during the early Universe
\citep{2003LNP...631..301C}. Massive stars with initial masses $\ga 20
\,M_{\odot}$ create sufficiently massive cores at the end of their
luminous lifetimes that they cannot avoid collapse to black holes
under self-gravity, although stars with masses above $\ga
100\,M_{\odot}$ at low metallicities may undergo pair-instability
supernovae and leave no remnant
\citep[e.g][]{2002ApJ...567..532H}. Massive stars that are below $\sim
20 \,M_{\odot}$ or those that experience a binary interaction instead
collapse to a neutron star remnant. This neutron star may then
collapse into a black hole as a result of mass transfer from a binary
companion. Predicting the rates of formation and merger of black-hole
systems thus requires stellar population synthesis: the process by
which stellar evolution models are combined and weighted according to
an initial mass function.  Such models have a long history
\citep[e.g.][]{TY73,1976ApJ...203...52T}.

Today there is growing evidence that most massive stars are in binary
or multiple systems, with 70 per cent of massive stars having their
evolution affected by binary interactions as shown by direct and
indirect observations \citep[e.g.][]{vanbev1998,EIT2008, Sana2012,
  Sana2014}. The presence of a nearby stellar companion can cause a
star to experience very different evolutionary pathways to those of
isolated stars. In general these complicate stellar evolution,
allowing extra opportunities for mass loss and mass gain.

There are several mature binary population synthesis codes world-wide
\citep[e.g.][]{vanbev1998,2002MNRAS.329..897H,IRT2004,2014MNRAS.440.1193L,Mennekens16}. Most
have been used to make predictions for the merger rates of compact
objects and the mass range for such mergers
\citep[e.g.][]{2007ARep...51..308B,2014MNRAS.440.1193L,deMinkBelczynski2015,Kowalska2015,Belczynski15,Mandel2016,MandeldeMink2016}.
With the detection of GW 150914 we now have a first observational
datum to compare against such models; a situation analogous to the
first detection of a supernova progenitor star (that of SN 1987A, now
known to be a rare progenitor type) in pre-explosion imaging which
provided an immediate constraint on stellar models
\citep{1987ApJ...321L..41W,1992PASP..104..717P}. While we should
exercise caution, since GW 150914 may prove similarly atypical of its
population, it is nonetheless useful to analyse the detection in the
light of theory.

In this article, we calculate the expected rate and parameters for
BH-BH mergers from our BPASS v2.0 models, and how these vary with
initial metallicity of the stellar population. In section
\ref{sec:method} we describe the BPASS stellar population synthesis
code and the numerical method employed in this analysis. In section
\ref{sec:results1} we present rates and timescales for binary black
hole mergers. We also identify a metallicity cut-off, indicating that
the progenitor system likely formed in a low-metallicity environment,
as predicted by \citet{Belczynski10}. In section \ref{sec:ratio} we
show that, at low metallicities, the rate of black hole mergers peak
for black hole binaries with near equal mass objects, similar to the
reported mass ratio of the progenitors of GW\,150914. Finally, in
section \ref{sec:follow-up} we discuss the implications for searching
for electromagnetic counterparts of such events, before presenting our
conclusions in section \ref{sec:conclusions}.

\vspace*{-10pt}
\section{Numerical Method}\label{sec:method}

\subsection{BPASS description and initial parameters}

The Binary Population and Spectral Synthesis, \textsc{BPASS}, code was
first discussed in \citet{EIT2008}, which also outlines modifications
to the Cambridge STARS code used to create the stellar evolution
models. A key difference between BPASS and most other codes is our use
of a large grid of 250,000 stellar models to follow the evolution of
interacting binary stars \citep[e.g.][]{vanbev1998}, rather than using
the approximation methods employed by rapid population synthesis codes
\citep[e.g.][]{2002MNRAS.329..897H}. The rapid method allows for the
uncertainties of binary evolution and their impact on predictions to
be explored; this would be too computationally intensive for detailed
stellar models. The use of detailed models, on the other hand, allow
us to accurately follow how the stellar envelope responds to mass loss
-- key to determining the eventual mass and fate of the star. The
spectral synthesis of stellar populations from individual stellar
models was described in \citet{ES2009,ES2012}, while a study of the
effect of supernova kicks on the stellar populations and supernovae
was described in \citet{ELT2011}. Many of the results for the code are
available at \texttt{http://bpass.auckland.ac.nz}.

BPASS models have been tested by the authors and others against
resolved and unresolved massive stellar populations in our Galaxy,
nearby galaxies and those at high redshift
\citep[e.g.][]{EIT2008,ES2009,ELT2011,ES2012,2014MNRAS.444.3466S}. They
have also been tested against directly detected SN progenitors and
relative SN rates
\citep[e.g.][]{2013MNRAS.436..774E,2015MNRAS.446.2689E,2015MNRAS.452.2597X}. Furthermore
we have recently released version 2.0 of BPASS \citep[][Eldridge et
  al., in prep.]{SEB2016}. This incorporates many refinements to the
code and its outputs compared to the earlier versions. The results of
BPASS v2.0 have already further demonstrated the improvement in
agreement between observations and stellar population models that
arises from the inclusion of interacting binaries
\citep[e.g. ][]{SEB2016,2016arXiv160103850W,2015arXiv151203214W,2016arXiv160107559M}.

While BPASS has been described in detail previously we provide a
detailed summary of BPASS here for those unacquainted with the
code. We use an initial-mass function (IMF) based on
\citet{1993MNRAS.262..545K}, with a power-law slope of -1.3 between
initial masses of 0.1 to 0.5\,M$_{\odot}$ and a slope of -2.35 from
0.5 to 300\,M$_{\odot}$. The stellar mass function therefore dictates
that less massive stars are more numerous and therefore fewer massive
remnants and black holes are generated than in a standard, unbroken
Salpeter IMF.  This is combined with an initial-mass ratio of
$M_2/M_1$ that is uniformly distributed between 0 to 1. All secondary
stars contribute to the stellar mass but we do not include a companion
in the total stellar mass estimate if its initial mass is less than
0.1\,M$_{\odot}$.

Key refinements in BPASS v2.0 \citep[relative to the v1.1 models
  discussed in][]{EIT2008,ES2009,ELT2011,ES2012} that affect the
results of this paper are as follows. First we increase the number of
models we have for our entire population from 15,000 detailed stellar
evolution models to 250,000 which represents several years of
computational time if run on a single processor. This increase allows
us to sample the initial parameter space for our initial masses at a
greater resolution. We have a grid of 68 initial primary masses from
$M_1=0.1$ to 300M$_{\odot}$, 9 values for the mass ratio, q, from
$M_2/M_1=0.1$ to 0.9 and 21 initial periods from 1 day to 10000 days
We also increase our grid of of initial metallicities to $Z=0.00001$,
0.0001, 0.001, 0.002, 0.003, 0.004, 0.006, 0.008, 0.010, 0.014, 0.020,
0.030 and 0.040.

The initial-period distribution is uniformly distributed in logarithm
of the period from 1 day to $10^4$ days. We note that by observing O
stars in the Galaxy, \citet{Sana2012} found that the observed period
distribution is somewhat steeper with a bias towards more close binary
systems. However \citet{2012ApJ...751....4K} found a flatter period
distribution in the Cygnus OB2 association that is consistent with
Opik's law, although their results did also suggest a slight
preference for short period systems. The uncertainty in assumed period
distribution is degenerate with uncertainties in the assumed model to
handle Roche-Lobe Overflow, Common-Envelope Evolution, tides and other
binary specific processes. Furthermore it is unknown whether the
observed period distributions should be extended to all stellar
masses. We can gain some insight into the effect of how varying the
initial period distribution will effect our results by looking at the
work of \citet{deMinkBelczynski2015} who find that the rates of
gravitational wave events will increase by a factor of two if a
distribution favouring short periods is used. Therefore we can say
that any predictions for our code are most likely a lower estimate on the possible
rate. 

We assume orbits are circular, or rather that the semi-latus rectum
distribution is flat. This can be assumed because, as shown by
\citet{2002MNRAS.329..897H}, the outcome of the interactions of systems
with the same semilatus rectum is almost independent of
eccentricity. This is also equivalent of assuming that systems are
circularised before interactions by tides. We only include tides in
our evolution models when a star fills its Roche-Lobe. Then we assume
tidal forces are strong and the star's rotation quickly synchronises
with the orbit. This is of course approximate and there are recent
studies have begun to explore how mass transfer may be different in
eccentric systems \citep{BDC15,DK16a,DK16b}. However we note that the
BPASS models have been tested to see if they can reproduce an observed
binary system with a slight eccentricity even after mass transfer
\citep{E2009}.
 
We scale the mass-loss rates applied from those observed in the local
universe, such that
$\dot{M}(Z)=\dot{M}(Z_{\odot})(Z/Z_{\odot})^{\alpha}$ and $\alpha=0.5$
\citep[except in the case of OB stars where $\alpha = 0.69$,
  see][]{Vinketal}. There is little consensus in the literature
regarding the definition of solar metallicity.
\citet{2014ApJ...787...13V}, for example, suggest the metal fraction
in the Sun is rather higher than usually assumed, while some authors
\citep{2002ApJ...573L.137A,2005ARAA..43..481A} suggest that Solar
metal abundances should be revised downwards to closer to Z = 0.014
\citep[also appropriate for massive stars within 500pc of the
  Sun,][]{2012AA...539A.143N}. We retain $Z_{\odot}=0.02$ for
consistency with the empirical mass-loss rates which were originally
scaled from this value. We note that at the lowest metallicities of
our models the small uncertainty in where we scale the mass-loss rates
will cause only small changes in the mass-loss rates due to stellar
winds. At the lowest metallicities mass-loss is primarily driven by
binary interactions.

A key feature of the BPASS models that sets them apart from others,
except the Brussels code \citep{vanbev1998,Mennekens16}, is that all
the interacting binary evolution models are evolved in a full
detailed stellar evolution code that is based on the Cambridge STARS
code and described in detail in \citet{EIT2008}. This greatly
increases our computational needs with the stellar models each taking
several minutes to calculate rather than fractions of a second. The
v2.0 models in this paper represent a total computing time on a single
computer of over 5 years. However while we have a computational cost
and therefore have to make assumptions, such as circular orbits, we
have significant gain in the accuracy of the stellar evolution
models. We find differences in how the stellar envelope responds to
mass loss relative to rapid population synthesis, as discussed in
\citet{EIT2008}. A comparison between our models and those of a rapid
population synthesis code show that our models would explode as red or
yellow supergiants while a rapid code assumed they would become
Wolf-Rayet stars.

We note we only compute one star in detail at a time. This is because
stars of very different masses have different evolutionary
timescales. Therefore computational time would be wasted on
calculating the evolution of a 1M$_{\odot}$ secondary star at the same
time as a 10M$_{\odot}$ primary. We therefore calculate the primary
evolution first, using the single star rapid evolution equations of
\citet{2002MNRAS.329..897H} to approximate the secondary's
evolution. We then recalculate the secondary's evolution in the
same detailed code either as a single star or in a binary with a
compact remnant depending on whether the binary is bound or
unbound. We also do not interpolate between these detailed models due
to the non-linear nature of binary evolution. The entire scheme is
discussed in greater detail in \citet{EIT2008} and \citet{ELT2011}.

One further refinement is vital for this work, that is the treatment
of the secondary models. In most supernovae, a binary is unbound in the
first supernova and thus the secondary evolves afterwards as a single
star. However in the case of those that remain bound BPASS selects
from a grid of binary models where the secondary is a compact remnant
to represent the further evolution of what was originally the
secondary star. Due to computational constraints in \citet{EIT2008} we
only assumed three masses for compact remnants, 0.6, 1.4 and
3M$_{\odot}$ for white dwarfs, neutron stars and black holes. In BPASS
v2.0 we now calculate the full range of possible secondary evolution
in binaries, allowing for a range of masses for the compact remnant
from 0.1 to 300M$_{\odot}$. We stress that this is different to rapid
codes that would typically take the evolutionary outcome of the
secondary and continue to evolve it. Due to our use of detailed
stellar evolution models this is still not computationally
feasible. Our extended selection of secondary models allow us to
follow evolution up to the formation of massive double black hole
binaries.

We follow the evolution and include mass-transfer when a star fills
its Roche-Lobe, full details are given \citet{EIT2008}. We assume any
mass lost from the primary is transferred to the secondary but this
can only be accreted by the secondary on a thermal timescale, any
extra mass is lost from the system. If the star filling its
Roche-Lobe engulfs the other star we assume common-envelope evolution
occurs. In a detailed stellar evolution code we cannot simply remove
the stellar envelope instantaneously \citep[as
  described in][]{EIT2008}. Instead we increase the mass-loss rate to
as high as numerically possible. Then at each timestep we calculate
the binding energy of material lost in our common-envelope wind and
remove this from the orbital energy of the star's core and the
secondary star. If the secondary star fills its Roche-Lobe we merge
the two stars together. This is, of course, approximate and different to
the typical implementation in other models but a reasonable
approximation. It is likely to be one of the sources of differences
between our models and those of others.  A consequence of our method
is that it is not straightforward to vary the physics of binary
interactions, since this would require recalculation of the entire
model set. While we do explore some key parameters, such as IMF, with
different model grids, we do not vary others, such as those controlling
common envelope evolution.

\subsection{Quasi-homogeneous evolution}

Our models only account for rotationally induced mixing in a simple
way. We assume that if a secondary star in one of our models accretes
more than 5 per cent of its initial mass and it is more massive than
2\,M$_{\odot}$ then the star is spun-up to critical rotation and is
rejuvenated due to strong rotational mixing. That is, it evolves from
the time of mass transfer as a zero-age main sequence star \citep[this is
similar to the method used by][]{vanbev1998}. At high
metallicities we assume that the star quickly spins down by losing
angular momentum it its wind so that there are no further consequences
to evolution. However with weaker winds at lower metallicity,
$Z\le0.004$, we assume that the spin down occurs less rapidly and so
the star is fully mixed on the main-sequence and burns all its
hydrogen to helium. We assume the star must have an effective initial
mass after accretion $>20$\,M$_{\odot}$ for this evolution for
occur. These limits were taken from the work of
\citet{2006A&A...460..199Y}. We note that we have discussed the
importance of these quasi-homogeneously evolving (QHE)
stars in greater detail in \citet{ELT2011,ES2012} and \citet{SEB2016}
showing there is strong observational evidence they exist. The key
difference we find from QHE changing the formation of BH-BH binaries
in our models is that it greatly increases the chance that the second
black hole to form is the more massive remnant. Because a QHE star
never evolves to the RSG phase the mass loss it experiences is less
and more of the mass accreted by the secondary is retained during the
star's evolution.

Furthermore our QHE is the result of mass transfer
only. \citet{MandeldeMink2016} and \citet{2016A&A...588A..50M} invoke
a different mechanism to create QHE stars. They consider the closest
binaries with the shortest rotation periods, of the order of a day, so
that both stars experience QHE with the rapid rotation induced due to
tidal interactions between the stars. While this is a very plausible
pathway, we have not yet included this in our standard BPASS evolution
models. Here we concentrate on the mass-transfer pathway to QHE. The
rates we predict here might still be further boosted if we were to
include such additional pathways. At the current time in our
population the binaries in \citet{MandeldeMink2016} and
\citet{2016A&A...588A..50M} are likely to result in mergers or
Roche-Lobe overflow because we do not assume those stars experience
QHE due to tidal forces.

Finally we note that we assume QHE is due to rotational mixing and it
different from the type of chemically homogeneous evolution found in
models of very massive stars, $\ga150M_{\odot}$.
\citet{2013MNRAS.433.1114Y} find such stars have very large convective
cores so that mass-loss can expose material from the core rapidly
without the need for rotational mixing.

\subsection{Predicting the remnant masses and SN kick velocities}

To estimate remnant masses our stellar models produce we use the
method described in \citet{ET2004}. We calculate how the binding
energy of the star varies with stellar radius. We assign as ejecta all
material that is above the point where the binding energy is equal to
$10^{51}$\,ergs. We assume a black hole is formed when the remnant
mass is above 3M$_{\odot}$, otherwise we assume a neutron star forms
with a mass of 1.4M$_{\odot}$. While this method is approximate, it
does give a a reasonable estimate for the size of the remnant and
agrees with other predictions for the initial masses when neutron
stars or black holes are formed during core-collapse
\citep[e.g.][]{2003ApJ...591..288H,2012ApJ...757...69U,2015arXiv151004643S,2015MNRAS.451.4086S}. We
note that the link between final remnant mass and initial mass can be
highly non-linear. For example, stars above $\approx 20$M$_{\odot}$
form black holes, while stars below form neutron stars. However at
higher initial masses, mass-loss can still lead to the formation of
neutron stars
\citep[e.g.][]{ET2004,2003ApJ...591..288H,2012ApJ...757...69U,2015arXiv151004643S}.
We do not include remnants from stars that end their evolution with
helium core masses between 64 to 133M$_{\odot}$ which are thought to
explode in pair-instability SNe and leave no remnant
\citep{2002ApJ...567..532H}.

Once the remnant and its mass is determined we use the kick
distribution of \citet{2005MNRAS.360..974H} to pick a kick velocity
and direction at random. For black hole kicks we assume a momentum
distribution and reduce the kick velocity by multiplying it by
1.4M$_{\odot}$ and dividing by the black hole mass. We do this because
of the growing evidence as discussed by \citet{Mandel2016} that
black-hole kicks are smaller than those of neutron stars. The full
method of determining kicks and the fate of the binary when a SN
occurs within it are described in \citet{ELT2011}.

\begin{figure}
\includegraphics[angle=0, width=0.48\textwidth]{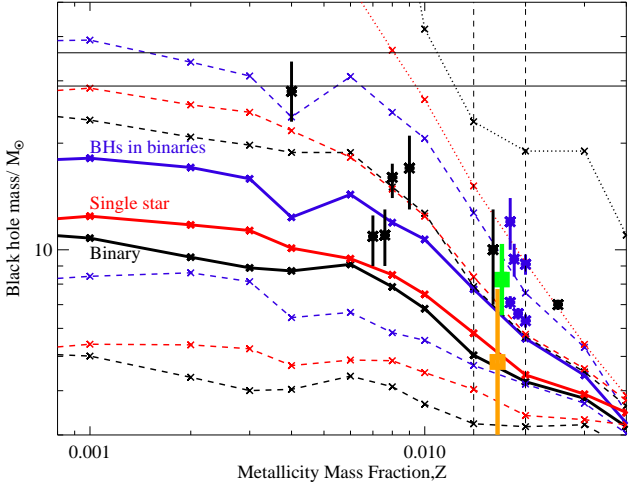}
\caption{The mean and maximum black hole natal masses from our
  models. The solid line is the mean black hole mass, the dashed lines
  indicate the 1$\sigma$ ranges while the dotted line is the maximum
  black hole mass. The red lines are for single star models and the
  black lines are for our binary models. The blue line represents the
  mean mass of black holes that remain in binary systems. The solid
  thick vertical black lines with asterisks represent the black hole
  masses collated in \citet{2010MNRAS.403L..41C}. While the blue
  asterisk and lines represent the Galactic black hole masses from
  \citet{2010ApJ...725.1918O}. Since the metallicities of Galactic
  black hole progenitors are unconstrained, we show them as lying
  close to Z=0.02, with small offsets for clarity. Vertical lines
  indicate two widely used values for Solar metallicity. We also plot
  the mean mass of these Galactic black-hole binaries in the green
  square and the mean mass of the single black-hole candidates
  identified from gravitational microlensing by \citet{microlense} in
  the orange square. The two thin horizontal lines indicate the masses
  of the black holes from GW150914.}
\label{figbhmass}
\end{figure}

To provide some test of the accuracy of this estimation we have
compared our predicted black hole masses to those observed in nature
in Figure \ref{figbhmass}. We see that the black hole masses predicted
by our single star and binary star populations are similar and agree
with the increasing trend of black hole masses with metallicity from
the results of \citet{2010MNRAS.403L..41C}. There have also been some
suggestions that there is a gap in the black hole masses expected from
stellar evolution, i.e. that there are no black holes in the mass range
between 3 to 5M$_{\odot}$ \citep{2012ApJ...757...91B}. We see in
Figure \ref{figbhmass} that fewer than half of all black holes should
have masses in this range in our Galaxy. Also binary systems
containing these objects are more likely to become unbound as they
will have larger kicks in our population under our assumptions for
natal black hole kicks. We show in Figure \ref{figbhmass} the mean
mass of black holes in binary systems. This is higher than the overall
mean and the observed black holes do lie closer to this line. Future
observing campaigns for mergers involving black holes will show if
there are such systems. We note here that we may over predict the BH-BH
merger rate at higher metallicities because we include these objects.

Recently \citet{microlense} have detailed a number of candidate single
compact remnants in the Galaxy discovered via gravitational
microlensing. They found no evidence for a gap in the mass of remnants
between neutron stars and black holes as suggested by
\citet{2012ApJ...757...91B}. We have estimated a mean mass for
single/runaway black-holes from the sample of \citet{microlense}. We
have assumed the same minimum mass for an object to be a black hole as
in our models of 3M$_{\odot}$. The mean is then based on seven objects
with the most massive being 9.3$^{+8.7}_{-4.3}$M$_{\odot}$. There are a
further 3 objects in their sample between 2 and 3$M_{\odot}$,
including these would reduce to mean to 3.8M$_{\odot}$. We see that
this mean is close to the mean black hole mass predicted at Solar
metallicity for our model populations. Again the sample is small but
future events from gravitational microlensing will increase our
understanding of black hole formation as much as future gravitational
wave sources.

We further note that while the mean black hole mass from a binary
population is slightly less than that from a single star population,
our models show the maximum mass of a black hole is greater from our
binary populations. This is because mergers and mass transfer in our
models allow stars to regain some of their lost mass from a companion
and also to create stars more massive than our assumed 300M$_{\odot}$
upper mass limit, although all binary evolution codes include this
possibility. Also all codes include the mass of the black holes
being able to increase via accretion from their companion star
\citet[e.g.][]{2015ApJ...810...20W}.

We also find that, while \citet{Belczynski16} suggest binary
interactions only lead to less massive black holes in a population for
the specific case of forming close massive black-hole binaries we find
this situation is not that simple. We find that some interacting
binaries actually lead to a more massive black hole than possible for
a star of the same mass from single-star evolution. We find that in
some cases a binary interaction during or before first dredge-up
(i.e., when the star becomes a red supergiant) in a model can prevent
or weaken dredge-up. This leads to a more massive helium core for the
stellar model than expected if it was single, and therefore eventually
a more massive black hole. This increase of the final black hole mass
is of the order of 10 per cent. We believe this difference is due to
our use of a detailed stellar evolution code and thus the ability to
follow how the development and extent of convective zones within the
stellar interior are affected by a binary interaction. Without this,
so subtle an effect might be missed.

\subsection{Compact remnant merger time calculation}

Our only addition to the BPASS code specifically predict the
merger rate of black-hole binaries has been to use the final orbital
parameters after the second supernova in the binary system to
calculate how long it will take for the two black-holes to merge. In
common with \citet{MandeldeMink2016}, we use the analytic form of
\citet{1964PhRv..136.1224P} to calculate this merger time. Peters gives
two limits for the merger time for low and high eccentricity
orbits. For rapid computation of our population we interpolate between
these two limits linearly. While a full solution for evolution over
time might give a more precise solution, the uncertainties introduced
by interpolation are smaller than those implicit in the assumptions
and uncertainties involved in other aspects of stellar population
synthesis.

\begin{figure*}
\includegraphics[angle=0, width=\textwidth]{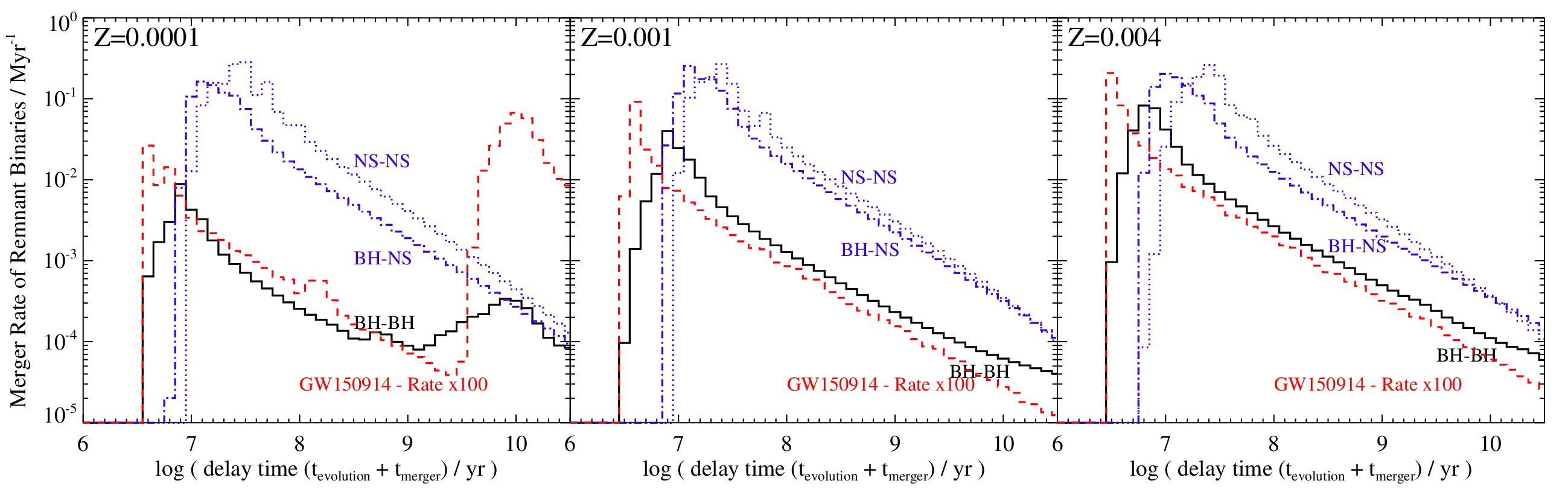}
\includegraphics[angle=0, width=\textwidth]{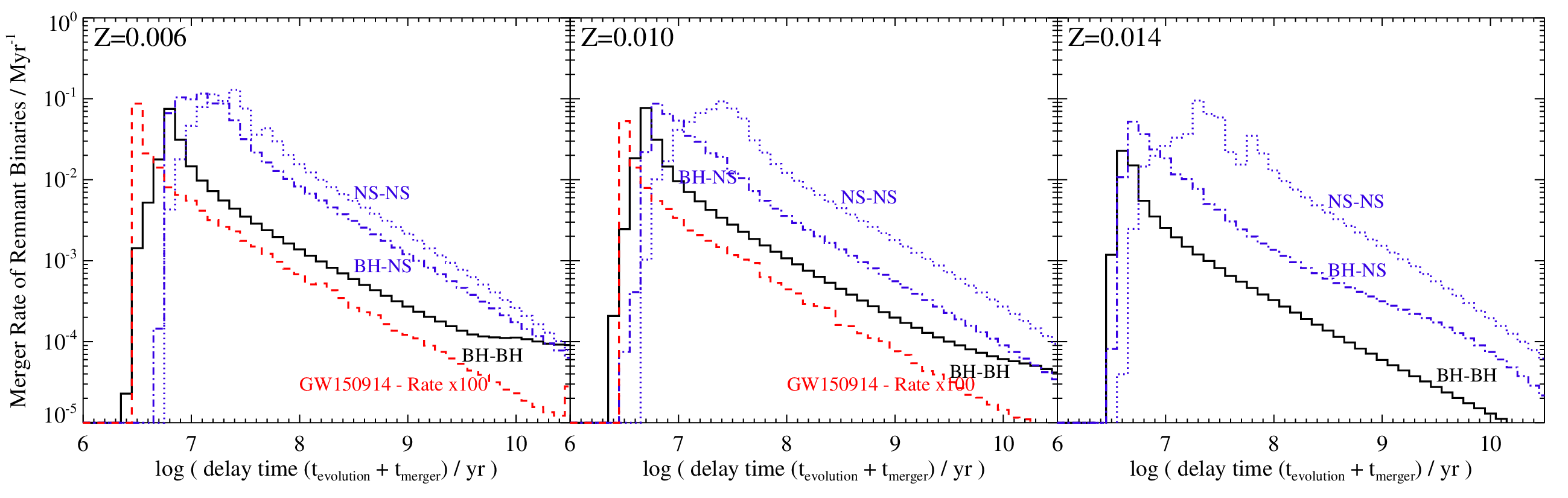}
\caption{The merger rate of systems with different delay times at
  various metallcities. They are calculated assuming an instanteous
  burst of star-formatino at $t=0$ with a total stellar mass of
  $10^6M_{\odot}$. In all cases the merger time dominates the delay
  time but the stellar lifetime does give rise to the minimum total
  time ($\sim$3\,Myr). The thick black line line shows the delay time
  distribution for BH-BH mergers. The red dashed line represent the
  population of systems matching GW\,150914 with the rate boosted by a
  factor of 100 for visibility.  The blue dotted line is the rate for
  NS-NS binaries and the blue dash-dotted line for NS-BH mergers.}
\label{figtimesdist}
\end{figure*}

To calculate the delay time for a black hole merger event we calculate
an evolution time, $t_{\rm evolution}$, defined as the interval after
the onset of star formation required for the progenitor stars to
evolve and create the two black holes. We combine these evolution
timescales with the time required for inspiral to calculate a total
required delay time, $t_{\rm delay}$. We then use this to calculate a
Galactic merger rate by assuming a constant star-formation rate of
3.5\,M$_{\odot}$\,yr$^{-1}$ for 10 Gyrs, i.e. we predict the number of
black-hole binary mergers expected per year if the Milky Way was made
gradually, of stars with a single metallicity. This prescription for
Galactic rate has previously been used by \citet{Dominik2013} and
\citet{deMinkBelczynski2015}.

For later use, we also estimate the rate of binary black-hole mergers
which have two black-holes in the mass ranges inferred for the
progenitors of GW\,150914. We do this by taking the number of mergers
from the bin element used for Figure \ref{figmasses} (see later) that
the masses of the GW\,150914 system lie in. This is all mergers with
the masses of both black holes individually between 23.7 and
42.2M$_{\odot}$.

We show our delay time distributions for sample metallicities in
Figure \ref{figtimesdist} for NS-NS, BH-NS and BH-BH mergers. The
merger rates are present assuming an instanteous burst of
star-formation at $t=0$ with a total stellar mass of
$10^6M_{\odot}$. We also indicate the delay time distribution for
GW\,150914 type events. We see that for all distributions there are a
short timescale peak and a longer plateau. We discuss other aspects of
the distribution below.

\vspace*{-10pt}
\section{Binary black hole merger timescales and rates}\label{sec:results1}

\begin{figure}
\includegraphics[angle=0, width=0.48\textwidth]{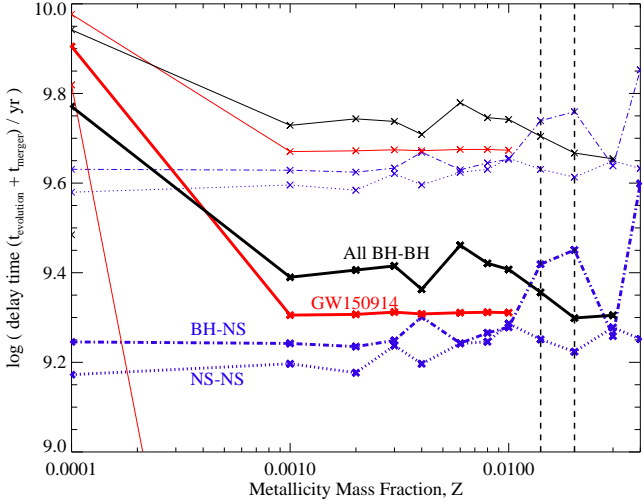}
\caption{The mean total time to merge (i.e. $t_{\rm evolution}+t_{\rm
    Merger}$) as a function of metallicity. In all cases the merger
  time dominates the delay time but the stellar lifetime does give
  rise to the minimum total time as shown in Figure
  \ref{figtimesdist}. The thick black line line shows the mean
  timescales for the full black-hole binary merger population with
  1\,$\sigma$ uncertainty indicated by the thin solid line.  The red
  lines indicate the equivalent for systems with stellar masses
  matching those of GW\,150914. Vertical lines indicate two widely
  used values for Solar metallicity. The blue dotted lines are the
  merger times for NS-NS binaries and the dash-dotted the time for
  NS-BH mergers.}
\label{figtimes}
\end{figure}

Figure \ref{figtimes} presents the variation in the total lifetimes
for mergers of binary black holes with metallicity. Our results are
consistent with the results of \citet{2012ApJ...759...52D} in that the
typical delay times are a few Gyr although there is a broad range of
possible merger times as shown in Figure \ref{figtimesdist}. We
identify cases where the merger occurs after a much shorter delay due
to the kick during the second core-collapse event reducing the orbital
period or inducing a reasonable eccentricity that shortens the merger
time. There are also cases with delay times longer than 10 Gyrs but we
do not include them in our rate estimate below, although we note that
some authors \citep[e.g][]{Belczynski16} include these. Delay times
are relatively independent of initial stellar metallicity, although
there is a slight trend to fewer mergers at higher metallicities. This
is due to more mass loss from the systems widening the orbit the
progenitor binary systems as well as the reduction in the typical mass
of the black holes.

At the lowest metallicities of $Z\le0.0001$ we see there is a trend
for it to be more common for black holes to have longer merger
times. The reduced opacities at this metallicity cause stars to be
more compact and so we find common-envelope evolution is less likely
and more mass can be transferred in binary interactions. This leads to
more systems experiencing QHE with relatively wide orbits. When these
form binary black hole systems, most mass goes into the black hole so
such systems have low eccentricities and take a longer time to
merge. We see, for example, that if the progenitor of GW\,150914 had
this metallicity it is most likely to have had a long merger time
rather than being a prompt merger. We note that \citet{Belczynski16}
also find the systems are dominated by long merger times at low
metallicity and see similar evolution even without including any QHE
in their models.

While the rate of mergers from these low metallicity progenitors is
higher than that at near-Solar metallicities, the number of stars that
formed at this metallicity is uncertain. We note that even by
$z\sim7-8$ there is strong evidence for much higher metallicities in
the intergalactic medium \citep[e.g.][and references
  therein]{2013ApJ...772...93K} and above this redshift the cosmic
star formation density history is very low
\citep[e.g][]{2014ARA&A..52..415M}, so the fraction of the cosmic star
formation history which takes place at such low metallicities is
probably very small.

\begin{figure}
\includegraphics[angle=0, width=0.48\textwidth]{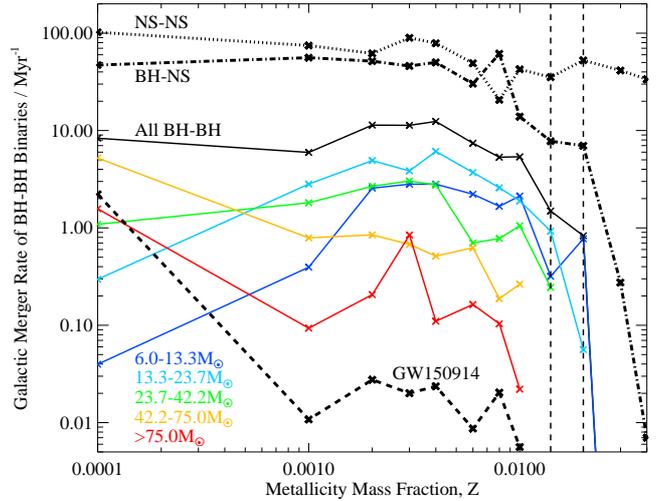}
\caption{The Galactic merger rate of black-hole binaries, given a
  constant star-formation rate of 3.5\,M$_{\odot}$\,yr$^{-1}$ for
  10\,Gyrs, as a function metallicity and the total mass of both black
  holes in the binary. The solid-black line is the total Galactic rate
  of all the black-hole merger population while the dashed-black line
  is for systems matching GW\,150914. The colour lines are the rates
  for logarithmically spaced mass bins of the mean black hole mass in
  the binary with ranges $\pm 0.25$\,dex. Note the structure in
  metallicity behaviour. The dotted and dash-dotted lines are the
  rates for NS-NS and NS-BH binaries.}
\label{figrates}
\end{figure}

Figure \ref{figrates} and Table \ref{qheratio} demonstrates that our
estimated Galactic merger rates (as defined in section
\ref{sec:results1}) are more metallicity dependent than those reported
by \citet{2012ApJ...759...52D}. The approximate range of rates are of
the same order as those given by \citet{deMinkBelczynski2015}. At
super-Solar metallicities the black-hole merger rate plummets: stellar
mass loss during the giant phase causes a decrease in the formation
rate and typical mass of black-hole binaries and the systems are much
wider at formation and fewer in number. The same trend can be seen in
the neutron-star/black-hole binaries where the rate also
decreases. Only the double neutron-star merger rate remains relatively
constant. We note that these rates are those derived for systems that
merge within 10Gyrs; it is likely that some systems at high
metallicities may still merge, but on timescales exceeding this
cut-off.

Below Solar metallicity, mass loss becomes less efficient and our
predicted binary black hole merger rate increases slightly before
plateauing at metallicities of $Z=0.010$ and below. We also consider
the relative rates in logarithmically spaced mass bins. The overall
merger rate is dominated by binary systems with a total black-hole mass
in the range 13-24\,M$_{\odot}$. At lower metallicities, a slight increase in
total rate is driven by a larger number of more massive systems
(24-42\,M$_{\odot}$), while the merger rate of less massive systems (6-14\,M$_{\odot}$)
declines. Very few black hole systems with total mass of 100\,M$_{\odot}$ are
generated in our models, with small number statistics in each
metallicity leading to variable rate estimates with a large associated
uncertainty. These represented a small contribution to the total rate
at all but the lowest metallicities.

When QHE is included at $Z\le0.004$ there are changes to the
rates. The main change is that at $Z=0.004$ it causes the mean delay
time to decrease by 0.1~dex, as seen in Figure
\ref{figtimes}. Otherwise the mean black hole delay time increases by
0.1~dex in our models over the range of metallicities while the other
merger times remain mostly constant. We find this is due to our
increasing black hole mass giving rise to smaller natal kicks, giving
rise to more circular black hole binaries being formed. Less eccentric
binaries have longer merger times giving rise to the shifting
distribution. At our lowest metallicities $Z\le 10^{-4}$ there is a
significant increase of merger times of 0.5 dex. This is due to more
of the mergers being dominated by QHE system which remain wider and
therefore require more time to merge.


We note that an alternative approach is to discuss predictions in
terms of event rate within the local universe, per cubic Gpc per
year. To obtain a meaningful estimate requires perfect knowledge of
the luminosity function, star formation history, current star
formation rate and metallicity distribution of galaxies in a given
volume - all of which are subject to considerable
uncertainties. However it is possible to make an order of magnitude
estimate for comparison with previous predictions. To do this we
assume a number of Milky Way equivalent galaxies within a cubic Gpc of
0.01~Mpc$^{-3}$ \citep[see, for example,][]{2010CQGra..27q3001A}. This
can be combined with our rates of events per Myr in a Milky Way-like
galaxy to produce a conversion factor of 1 Myr$^{-1}$ (Galactic)
$\approx$10 Gpc$^{-3}$~yr$^{-1}$ (volume averaged). Our predicted
merger rates are, very approximately for all black hole mergers, in the range of 10 to
100~Gpc$^{-3}$~yr$^{-1}$. This is comparable to the merger rate inferred
by \citet{2016arXiv160203842A} of between 2 and
400~Gpc$^{-3}$~yr$^{-1}$.

We note that in Figure \ref{figrates} our mergers of black hole
binaries with a total mass above 75M$_{\odot}$ have a discontinuity
between metallicities of $Z=0.002$ and 0.004. Below this metallicity
jump, stars that previously would have formed black holes experience
pair-instability SNe and so leave no remnant decreasing the rate of
mergers from the most massive binaries. The rate increases again at
the lowest metallicity with the increasing contribution from systems
that experience QHE as well as now having more stars that are massive
enough to avoid a pair-instability SNe and form very massive black
holes. There is also similar non-monotonic variation with metallicity
in some of the other merger rates. This is the result both of us using
detailed models over a finite grid of masses and also that theoretical
models suggest there is not a simple mapping between initial masses
and remnant mass
\citep{2003ApJ...591..288H,2012ApJ...757...69U,2015arXiv151004643S},
which also changes with metallicity \citep[see Figure 6
  in][]{ET2004}. There is some observational evidence for this from
the fact that magnetars (highly magnetic neutron stars) arise from
stellar progenitors with initial masses above $40M_{\odot}$ as well as
lower mass progenitors of $\approx 17M_{\odot}$
\citep{2008ApJ...685..400B,2009ApJ...707..844D}. These studies
suggest that the final remnant masses are likely to be determined by
mass-loss by winds and binary interactions but also the detailed
nature of the final stages of nuclear burning.

The point to point variation in behaviour gives some
indication of the uncertainties from our adopted methods, and are of
the order of 0.3 dex at most. Our event rates for neutron-star
mergers, for example, is sensitive as well to the minimum mass for
core-collapse SNe. This is 8M$_{\odot}$ at $Z=0.020$ but decreases to
6M$_{\odot}$ at the lowest metallicities; at the same time the number
of black holes forming is also changing and so the rate appears to to
jump.

In Table \ref{qheratio} we show the typical orbital parameters for the
binary black hole systems that merge in our simulations. We see that
in general the systems require a high eccentricity to merge within
10Gyrs. This validates our earlier assumption of interpolating between
the limits of high and low eccentricity derived by
\citet{1964PhRv..136.1224P} as most systems are highly eccentric after
the second SN. Furthermore we note again that at the lowest
metallicities there is a significant decrease in the mean
eccentricity. This is due to the fact that at lower metallicities the
QHE stars are significantly more compact at the end of the evolution
and eject very little mass in the second supernova. Therefore the
orbit remains relatively circular and so the merger time remains high.

We see a similar pattern for the typical system giving rise to events
matching GW150914 in Table \ref{typical}, although due to the
restriction of the masses we find that at higher metallicities the
possible systems are very few. At lower metallicities there are two
ranges of initial masses due to pair-instability SNe preventing stars
from forming black holes at some mass ranges. Again we see that most
of our systems are required to have high initial eccentric when
formed, unless the systems are at the lowest metallicities when QHE is
the dominant formation channel - although as we see in Figure
\ref{figtimesdist} these have long merger times. Since these would
have formed earlier in the Universe at these metallicities it is
possible that this is a likely formation channel for
GW\,150914, although as we discuss above few stars were formed at
low metallicities.

Finally we provide a very approximate estimate of how detectable the
different mergers will be in Table \ref{chirp}. For each metallicity
we calculate the mean chirp masses for all our compact remnant mergers
in our calculations where, $\mathcal{M}_0= (M_1 M_2)^{3/5} /
(M_1+M_2)^{1/5} $. The distance at which a merger can be detected is
related to this value by $d \propto \mathcal{M}_0^{5/6}$, the volume
within which mergers can be detected is thus $\propto
\mathcal{M}_0^{5/2}$. Rather than calculate an absolute rate we
calculate the rates relative to the mean merger rate for all NS-NS
mergers. We see that, except at the lowest metallicity, the NS-NS
merger rates are similar. However with the higher chirp masses for
systems involving a black hole our approximate detection rate increase
significantly. This echoes the predictions of
\citep{Belczynski10,Dominik2015} that a binary black-hole would be the
first remnant mergers to be detected.

\begin{table*}

\caption{Fraction of mergers that arise due to QHE, for NS-NS, NS-BH,
  BH-BH and GW150914-like mergers. Also the Galactic merger rate for
  these same mergers and the typical eccentricities, $e$, initial
  BH-BH binary orbital separation, total BH binary mass and initial
  orbital separation for the BH-BH binaries.}
\label{qheratio}
\begin{tabular}{ccccccccccccccc}
\hline
            &   \multicolumn{4}{c}{Fraction of QHE systems}  &  \multicolumn{4}{c}{Galactic Merger Rate / Myr$^{-1}$}  &   &  $M_{\rm BH tot}$ &  $\log(P/$\\
$Z$       &    NS-NS  &  BH-NS   &   BH-BH  &  GW150914& NS-NS  &  BH-NS &BH-BH  & GW150914 &$e$  & $/M_{\odot}$ & days) \\
\hline
$10^{-5}$  &  0  &   0.061 &    0.878 &   0.989  &  160 & 29  & 3.1  & 0.14 &  0.34$\pm$0.32     &   72$\pm$49  &  0.7$\pm$0.7 \\
$10^{-4}$  &  0  &   0.008 &    0.858 &   0.988  &  100 & 47  & 8.3  & 2.2 &  0.24$\pm$0.31   &      67$\pm$36  &  0.7$\pm$0.6\\
0.001     &  0  &   0.011  &   0.721 &   0.000  &   75  & 56 & 6.0  & 0.011 & 0.92$\pm$0.16       &  28$\pm$15 &   1.7$\pm$0.9\\
0.002     &  0  &   0.023  &   0.692 &   0.000  &   62  &  52  & 11 & 0.028  &0.91$\pm$0.19     &   24$\pm$19 &    1.5$\pm$0.8\\
0.003     &  0.024& 0.026  &   0.653 &   0.0002 &   89  & 46  & 11  & 0.021 & 0.86$\pm$0.27     &   29$\pm$29  & 1.4$\pm$0.8\\
0.004     &  0.033&  0.024 &   0.685 &   0.049  &   79  &  50  & 12 &0.024 & 0.93$\pm$0.14    &    21$\pm$13  &  1.5$\pm$0.8\\
0.006     &  0   &    0    &     0   &     0     &  49 & 30 &7.4 & 0.009 &  0.84$\pm$0.26  &    21$\pm$16  &  1.2$\pm$0.8\\
0.008     &  0   &    0    &     0   &     0     &  21  & 62  & 5.3 & 0.019 &   0.89$\pm$0.16   &   21$\pm$20  &   1.2$\pm$0.9\\
0.010     &  0   &    0    &     0   &     0     &  43  & 14  & 5.4 & 0.006 &    0.87$\pm$0.22   &    20$\pm$12  &  1.4$\pm$0.8\\
0.014     &  0   &    0    &     0   &     0     &  35 & 7.8  & 1.5  & 0    &    0.95$\pm$0.11 &     17$\pm$5  &  1.8$\pm$0.7\\ 
0.020     &  0   &    0    &     0   &     0     &  52 & 7.0  & 0.82  & 0 &   0.98$\pm$0.02   &  10$\pm$2 &    1.6$\pm$0.5\\
0.030     &  0   &    0    &     0   &     0     &  41  & 0.27 & 2$\times10^{-7}$ & 0&  0.9996$\pm$0.0003     &    8       &   3.6$\pm$0.1  \\
0.040     &  0   &    0    &     0   &     0     &  34  & 0.007 &   0   &   0   &   0  &  0  &  0 \\
\hline
\end{tabular}
\end{table*}

\begin{table*}

  \caption{The parameters of systems that would result in
    GW\,150914-like progenitor binaries. We show mass ranges and
    typical initial periods for the stars at different phases of the
    binary progenitors lifetime. M$_{1,i}$, M$_{2,i}$ and $\log(P_{i,1}
    /{\rm days})$ are the initial masses and periods of the binary
    stars. M$_{1,BH}$, M$_{2,pSN}$ and $\log(P_{i,2} /{\rm days})$ are
    the typical mass of the black hole from the first SN, effective
    secondary mass range, including results of mass transfer and
    typical period post-SN. Finally M$_{1,BH}$ and M$_{2,BH}$ are the
    final masses of the black holes formed, we also show typical
    eccentricities, $e$, total BH binary mass and initial orbital
    separation for the BH-BH binaries.}
\label{typical}

{\hspace{-1cm}\small{
\begin{tabular}{ccccccccccccccc}
\hline
      &    M$_{1,i}$&  M$_{2,i}$& $\log(P_{i,1}$ &  M$_{1,\rm BH}$&  M$_{2,\rm pSN}$&  $\log(P_{i,2} $ &  M$_{1,\rm BH}$& M$_{2,\rm BH}$&  &  $M_{\rm BH tot}$ &  $\log(P/$\\
$Z$       &  /  M$_{\odot}$&  /  M$_{\odot}$& $ /{\rm days})$ & /  M$_{\odot}$&  /  M$_{\odot}$&  $ /{\rm days})$ &  /  M$_{\odot}$&  /  M$_{\odot}$& $e$   &/  M$_{\odot}$& days) \\
\hline
$10^{-5}$  &    40--80, &  20--90              &$\ge$0 &      25--40  &    35--100   & 0.6--0.8,  &     20--40  &  27--40                 &    0.05$\pm$0.08  &    79$\pm$4  &  0.7$\pm$0.2\\
          &     100     &                      &        &                &             &  $\ge$3.8   &            &           &                     &                      &       \\
$10^{-4}$  &    60--80, &  24--65             & $\ge$0.6&       25--40  &    40--70    &     $\ge$3.6       &     25--40  &  24--40       &   0.07$\pm$0.06  &     69$\pm$6   &  0.6 $\pm$0.2\\
          &      120    &                 &           &               &                 &                  &                &             &                     &                       &   \\
0.001     &    80, 100 &  40--72               & $\ge$0.6&         32--40  &    70--100   &      $\ge$3.6       &     32--41  &  28--41   &    0.9994$\pm$0.0006 &  67$\pm$6  &  4.0$\pm$0.3\\
0.002     &    120 &  40--110             &$\ge$0.8&              32--40  &    70--100   &      $\ge$3.2       &     25--41  &  25--35    &   0.9994$\pm$0.0006 &  64 $\pm$6   &  4.0$\pm$0.3\\
0.003     &    100--200, 300&  60--180          &$\ge$0.8&        32--40  &    80--100   &      $\ge$3.4       &     32--40  &  24--31    &  0.9993$\pm$0.0006    &   63$\pm$6   &  4.0$\pm$0.4\\
0.004     &    120--200, 300&  75--180        &$\ge$1&          25--40  &    100--120  &      $\ge$3.2       &     25--40  &  27--38      &  0.9994$\pm$0.0006   &    62$\pm$7 &  4.1$\pm$0.4  \\
0.006     &    100--300&  70--150      &$\ge$0&                 32--40  &    120--150  &      $\ge$3.4       &     25--40  &  24--41      &   0.9994$\pm$0.0007 &  68$\pm$9  &  4.1$\pm$0.5\\
0.008     &    200&  180             &$\ge$1.4&                   25--32  &    120--200  &    $\ge$2.4   &     25--34  &  26--37          &  0.9994$\pm$0.0007   & 57$\pm$6  &  4.1$\pm$0.5\\
0.010     &       200  &    120    &1.2&                     16--25   &    120      &    $\ge$2   &     25--40  &   25                    &  0.9991$\pm$0.0008  &   50$\pm$1  &  3.8$\pm$0.4\\
\hline
\end{tabular}}}
\end{table*}

\begin{table*}
\caption{Mean chirp masses for different types of compact remnant
  mergers and the approximate relative detection rate this implies
  relative to the mean rate of neutron-star neutron star mergers. This
  is calculated by assuming the volume within which mergers can be
  detected is given by $\mathcal{M}_0^{5/2}$, the mean rate of NS-NS
  mergers is then calculated as the mean of all the metallicities considered.}
\label{chirp}
\begin{tabular}{ccccccccccccccc}
\hline
            &   \multicolumn{3}{c}{ Mean Chirp Mass, $\mathcal{M}_0$}  &  \multicolumn{3}{c}{Relative detection rate} \\
$Z$       &    NS-NS  &  BH-NS   &   BH-BH  &    NS-NS  &  BH-NS   &   BH-BH\\
\hline						
$10^{-5}$&	1.22	&	3.08$\pm$1.04	&	27.3$\pm$18.1	&	2.44	&	4.52	&	115	\\
$10^{-3}$&	1.22	&	3.15$\pm$0.93	&	25.5$\pm$12.9	&	1.58	&	7.80	&	258	\\
0.001	&	1.22	&	3.06$\pm$0.86	&	9.47$\pm$4.40	&	1.16	&	8.64	&	15.5	\\
0.002	&	1.22	&	2.93$\pm$0.82	&	8.77$\pm$7.18	&	0.96	&	7.13	&	24.5	\\
0.003	&	1.22	&	2.88$\pm$0.72	&	10.9$\pm$11.3	&	1.38	&	6.10	&	42.1	\\
0.004	&	1.22	&	2.66$\pm$0.63	&	7.79$\pm$4.30	&	1.22	&	5.48	&	19.9	\\
0.006	&	1.22	&	2.61$\pm$0.61	&	8.07$\pm$5.64	&	0.76	&	3.13	&	12.9	\\
0.008	&	1.22	&	3.99$\pm$1.79	&	7.18$\pm$3.64	&	0.32	&	18.4	&	6.94	\\
0.01	&	1.22	&	2.54$\pm$0.61	&	7.21$\pm$3.42	&	0.66	&	1.35	&	7.07	\\
0.014	&	1.22	&	2.55$\pm$0.96	&	6.45$\pm$1.60	&	0.55	&	0.77	&	1.49	\\
0.02	&	1.22	&	2.14$\pm$0.41	&	4.07$\pm$0.68	&	0.81	&	0.44	&	0.26	\\
0.03	&	1.22	&	2.52$\pm$0.47	&	3.29$\pm$0.00	&	0.64	&	0.03	&  $5\times10^{-8}$	\\
0.04	&	1.22	&	1.87$\pm$0.10	&	0.00$\pm$0.00	&	0.52	&	0.0003	&	--	\\
\hline
{ Mean } & 1.22 & 2.8  & 11& 1  & 4.9  & 42 \\
\hline 
\end{tabular}
\end{table*}

\vspace*{-10pt}
\section{Effects of BH-BH Mass Ratio}\label{sec:ratio}

\begin{figure*}
\includegraphics[angle=0,width=\textwidth]{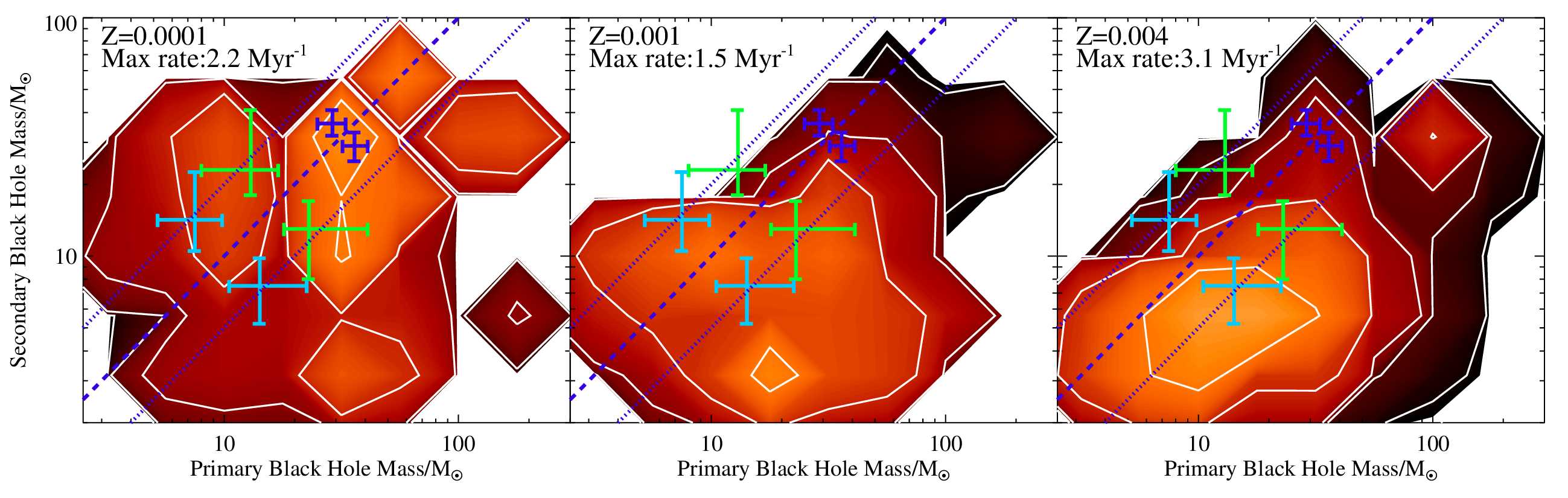}
\includegraphics[angle=0,width=\textwidth]{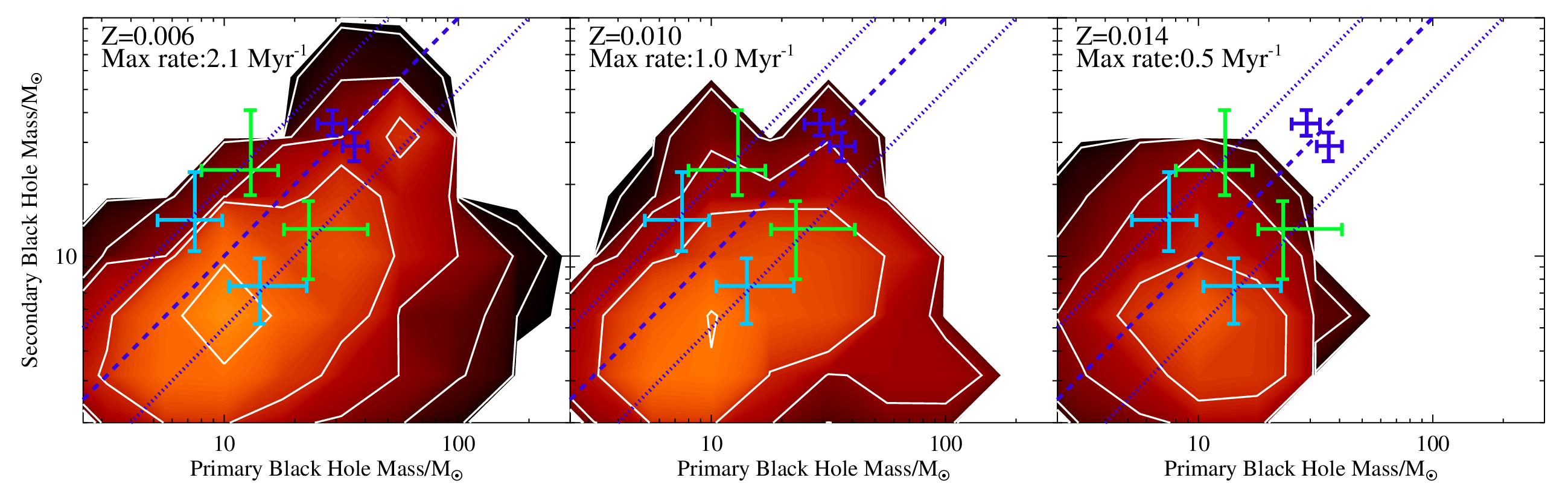}
\caption{The dependence of binary black-hole Galactic merger rate on
  black hole mass and mass ratio. The coloured contours represent the
  relative rates. The white contours are lines of constant merger rate
  for every order of magnitude, i.e. 1, 0.1, 0.01, 10$^{-3}$ and
  10$^{-4}$ \,Myr$^{-1}$ for a population forming stars at a constant
  rate of 3.5\,M$_{\odot}$\,yr$^{-1}$ over a 10\,Gyr period, we note
  in the $Z=0.014$ panel the maximum rate is too low for the first
  contour to be included. Blue crosses represent the inferred masses
  of the black holes in GW\,150914 and their uncertainties. The two
  points indicate different evolutionary behaviour dependent on
  whether the more massive black-hole was formed first or second. The
  dashed line represents a black hole mass ratio of unity while the
  dotted lines represent scenarios in which one black hole is twice
  the mass of the other. We also indicate the progenitor masses of the
  recently announced events GW\,151226 and LVT\,151012 \citep[][see
    section 6]{2016arXiv160604856T} in light blue and green
  respectively.}
\label{figmasses}
\end{figure*}

As Figure \ref{figmasses} illustrates, the rate of mergers at
$Z\ge0.004$ peaks in systems where the primary black hole is just over
twice that of the secondary (a mass ratio of 2), and is highest for
primary black holes around 10 M$_{\odot}$. Comparable rates (to within
an order of magnitude) are found for some binaries with mass ratios
ranging from 10 to 0.5, although this is dependent on both metallicity
and primary mass. At metallicities close to Solar, black hole binaries
with masses $>30$M$_{\odot}$ (in either the secondary or primary) are
rarely seen in our models. At lower metallicities ($Z<0.004$), the
structure in rates becomes more complex, with the highest rates
observed or systems in which the primary is still close to
10M$_{\odot}$ in mass but now comparable to, or even {\em less}
massive than, the secondary, although several regions of parameter
space with a range of mass ratios and primary masses show comparable
rates.  At the lowest metallicities we show ($Z=10^{-4}$), the rate
distribution of mergers is predicted to show two peaks of comparable
strength, one of which occurs at near-equal masses comparable to those
of GW 150914, while the other occurs at lower masses and more
asymmetric systems.

There are two competing pathways for the black hole mergers in our
synthetic population. The first is for systems that do not require
QHE. These are interacting massive binary systems driven together by
common-envelope evolution or wider binary systems that only experience
Roche-Lobe overflow and require a more eccentric black-hole binary to
merge quickly. For these systems typically the primary black hole is
the more massive one. Systems with a more massive secondary black hole
are less probable. For this pathway decreasing metallicity increases
the typical black hole masses slightly.

The second is an interaction leading to efficient mass transfer so
that the secondary star can accrete a large amount of material and at
low metallicities experience QHE. At $Z=0.004$ the peak from this
pathway overlaps with that from the typical mass-transfer. However as
the metallicity decreases the secondary star retains more of its mass
as stellar winds weaken and so the peak of events switches to the
secondary black hole being more massive. We show in Table
\ref{qheratio} the fraction of mergers at each metallicity that arise
from systems that have experienced QHE. We see this is a major channel
for most BH-BH mergers but, in agreement with \citet{Belczynski16} we
find that GW\,150914 is possible and more likely from interacting
binary evolution without QHE, except at the lowest metallicities we
consider. 

One difference between the standard and QHE pathways is that the QHE pathway
tends to have a longer delay time. In the standard binary pathway both
stars can interact, so when the second black hole is formed the stars
can already be in a tight orbit. For the QHE pathway the lack of
a second interaction leads to wider orbits and therefore longer merger
times. The extreme case of this can be seen in the $Z=0.0001$ delay
time panel of Figure \ref{figtimesdist} with the peak around
10~Gyr.

A final factor to consider is how many of our black hole mergers are
predicted to have a mass ratio close to unity. From summing the bins
with mass ratio consistent with the unity line in Figure \ref{figmasses}, we
find that at $Z\ge 0.001$ approximately 20 to 40 per cent of black hole
mergers should have similar black hole masses. At metallicities below
this value the number can drop to a less than 10 per cent because of
the QHE stars dominating the rate.

We note that this occurs in our observationally-benchmarked standard
model set, in the absence of fine tuning to match the GW\,150914
event. The observational benchmarking of BPASS has previously
concentrated on modelling observed stars, stellar systems and
core-collapse SNe. This is the first time we have confronted
predictions for compact remnants. GW150914 has provided a new test of
BPASS that other spectral synthesis codes such as
Starburst99 \citep[][]{1999ApJS..123....3L} and the
\citet{2003MNRAS.344.1000B} models are unable to pass due to their
reliance on single-star models.

\vspace*{-10pt}
\section{Implications for EM Follow-Up}\label{sec:follow-up}

While a low significance gamma ray burst transient coincident with the
GW 150914 event was reported by \citet{2016arXiv160203920C}, the
association of this with the binary inspiral and merger is unclear. In
a separate analysis \citet{2016arXiv160600314G} found this signal was
consistent with background fluctuations. Follow up with other
instrumentation ranging from radio to X-ray wavelengths
\citep{2016arXiv160208492A} failed to yield a detection. Such an
identification was always unlikely given the low luminosities
predicted for BH-BH merger counterparts, the poor sky localisation of
LIGO in its 2015 configuration, and the early detection at a time when
a number of the planned electromagnetic follow-up programmes
(e.g. BlackGEM, \citealt{2015ASPC..496..254B}, or GOTO,
http://goto-observatory.org/) were not yet on sky. BH-BH mergers are
considered poor candidates for electromagnetic detection, but this is
yet to be demonstrated observationally. As such, it is useful to
consider the implications of our analysis for hypothetical follow-up
of similar future events.

LIGO in its current configuration (two detectors in the United
States), can localise events to a region of order $\sim$500-1000\,deg$^2$,
with this region typically distributed in an arc with multiple high
probability regions rather than a single field
\citep{2015ApJ...800...81E,2014ApJ...795..105S}. Identification of
counterparts in such large regions is exceptionally difficult due both
to the small field of view of typical optical telescopes and the
number of `normal' electromagnetic transients expected per square
degree. Suggested follow-up strategies include optimised tiling of the
highest probability regions \citep[e.g.][]{2015arXiv151102673G} and
targeting the environs of the brightest galaxies in the region
\citep[which contribute $\sim$50 per cent of the stellar light,
  e.g.][]{2015arXiv150803608G,2011CQGra..28h5016W}.
While the former strategy is unaffected by our results, the effect on
the latter could be significant.  Galaxy catalogues compiled to date
have been based on a distance range and luminosity cut-off optimised
for neutron star-neutron star mergers, and tuned to select the typical
host galaxies of short GRBs \citep[e.g.][]{2015arXiv150803608G}. They
prioritise galaxies on B-band luminosity, focusing on the large scale
structures containing most luminous mass. Given the strong metallicity
dependence of our results, using such catalogues may not be a optimal
strategy for binary black hole mergers.

As discussed in section \ref{sec:results1}, we see two peaks in the
age distribution of GW\,150914-like progenitors at the lowest
metallicities we consider arising from different pathways. The first of
these allows systems to merge at ages of a few million years, while
the second dominates at ages of a few billion years upwards. At any
metallicity in our models above $Z=0.0001$, only the first of these pathways
is permitted.

While it is appealing to abandon the short-lived possibility in favour
of the longer lived progenitor pathway, there are potential
problems. The rate expected from long-lived progenitors in our models
peaks at 10\,Gyr and is only measurable at metallicities below a two
hundredth of Solar. In this scenario the stars that ended their lives
in GW\,150914 likely formed at $z\sim2$, and at metallicities
significantly lower than those estimated in the star forming galaxy
population at that redshift \citep[e.g.][]{2014MNRAS.445..794T}. They
would have to arise in pockets of pristine or near-pristine gas
isolated from pollution by nearby supernovae - very rare survivors of
the metal-poor Cosmic Dawn.  While it is possible to push the
formation redshift still earlier (i.e. invoking merger timescales
$>10$\,Gyr), we note that the star formation density at such early
times falls off rapidly was many orders of magnitude lower at an epoch
when the required metallicities were common \citep[see][for
  discussion]{2014ARA&A..52..415M}.

On the other hand, the short-lived progenitor pathway also requires
that the stars that ended their lives in GW\,150914 formed in a gas
cloud that is significantly below the volume averaged typical
metallicity at its redshift. One advantage is that such short-merger
systems form at a wide range of more moderate metallicities. At any
metallicity above $Z=0.001$, and given the event redshift,
$z\sim0.049$, the stars that ended their lives in GW\,150914 likely
formed at $z\le 0.4$, while more typical BH-BH merger progenitors
formed at $z\sim0.5$, an epoch at which the Universe was already
heavily metal enriched. However galaxies with metallicities below
half-Solar (required by our models) are are not unknown in the local
universe \citep[see e.g.][]{2013MNRAS.430.2097J}, although their
number density is low and remains poorly constrained.

Galaxies in the local Universe show a strong relation between
metallicity and both stellar mass and luminosity, with bright
($M_g\la-19$), massive (M$_\ast\ga10^{10}\,$M$_\odot$) galaxies
typically having Solar or super-Solar metallicities
\citep{2004ApJ...613..898T}. While there are likely to be pockets of
lower metallicity star formation in these systems, the bulk properties
mitigate against the formation of high mass black hole-black hole
binaries. Instead, short-timescale binary black hole merger events are
more likely to be associated with low mass, less luminous regions of
the cosmic web and potentially with post-starburst galaxies
(i.e. those which formed significant numbers of stars $\sim$3-5\,Gyr
ago).  Hence, even if galaxy catalogues were extended to the larger
distances expected for BH-BH mergers relative to NS-NS mergers, their
analysis of the densest regions of stellar material may still skew
observations away from the best candidates.

While we have not investigated NS-BH events in detail, we expect them
to exhibit a similar bias towards low metallicities - whether formed through
short or long timescale pathways.  We advocate the
addition of galaxy metallicity information to follow-up prioritisation
algorithms using galaxy catalogues. These should be employed with
care, particularly if the LIGO rapid analysis is able to constrain an
event as a likely binary black hole merger before electromagnetic
follow-up.

\vspace*{-10pt}
\section{Conclusions}\label{sec:conclusions}

We can conclude that the most likely evolutionary pathway for
GW\,150914 is standard binary evolution, in agreement with the
work of \citet{Belczynski16} who
independently reached the same conclusion. Having said that, there are
differences between these two population synthesis codes regarding
handling of core-collapse and the natal kicks of neutron stars and
black holes. This leads to our predictions giving wider black-hole
binaries that are highly eccentric and merge within 10~Gyrs, while
\citet{Belczynski16} have shorter period systems that undergo direct
black hole formation.

As more compact remnant mergers are detected there will be ever
tighter constraints on binary population synthesis. However we stress
this should only be one test and other observational data such as
stellar populations and supernovae should be used to constrain
population synthesis codes.  The results presented in this paper are
based on v2.0 of the BPASS models, which have already been tested
against observational constraints in a range of different areas from
the distant universe to local star forming galaxies \citep[e.g.][and
  references
  therein]{SEB2016,2016arXiv160103850W,2016arXiv160107559M}. Inevitably,
their results are somewhat dependent on the initial parameter
distributions employed by the models (for example in binary separation
distribution and mass loss rates), and we endeavour to use
observationally motivated constraints for all free
parameters. Importantly, however, we have not tuned our models to
achieve the results in this paper. We have simply taken the standard
BPASS model set and analysed its predictions for black-hole mergers.

We have calculated predictions for binary black hole merger timescales
and rates, and considered the 1st LIGO event detection in light of
these predictions. We find that the event must have come from a
metallicity of $Z=0.010$, roughly half Solar, or below and the rate
for such events is nearly constant for metallicities between 1/5th and
1/10th Solar via normally binary evolution at 0.1 to 0.4 per cent of
all binary BH mergers. This metallicity cutoff was estimated
independently by \citet{implications} using the single star models of
\citet{2015MNRAS.451.4086S}. However both these cutoffs are higher
than the metallicity cut-off given by the binary population
synthesis of \citet{Belczynski16} of 1/10th Solar or $Z=0.002$.  The
exact reason for this difference is not clear and it is likely to be a
combination of factors:
\begin{enumerate}
\item We may estimate more massive black holes in our simulations,
  either due to our method of estimating the remnant mass formed or
  our assumed stellar-wind mass-loss scheme or other model
  detail. However as shown in Figure \ref{figbhmass} our estimated
  black hole masses are not so high, and are consistent with
  observations of known Galactic systems.  If we were to change the
  physics to decrease the black hole masses then the observed
  black-hole masses in nearby stellar binaries would become extreme rather
  than typical systems.
\item Our black-hole kick model may be weak compared to others. We
  pick a kick at random from a Maxwell-Boltzmann distribution with
  $\sigma=265{\rm km\, s^{-1}}$ \citep{2005MNRAS.360..974H}, but using
  it as a momentum distribution for the black-holes so that the kick
  velocity is reduced by $M_{\rm BH}/(1.4M_{\odot})$. This
  means more massive black holes are likely to have
  weaker kicks and remain bound (see the blue line in Figure
  \ref{figbhmass}). This means more low-mass black holes will be
  runaway compact remnants and are more likely to be detected by
  gravitational microlensing \citep[e.g.][]{microlense}.
\item The detail of our treatment of common-envelope evolution could
  also produce more black-hole binaries. While many groups include
  common-envelope evolution it is the most uncertain phase of an
  interacting binary system \citep{IV13}.
\end{enumerate}
The differences between the different model populations are
essentially linked to the formation of black holes, both in how
massive they can be upon formation and what kick they obtain in their
birth. Future gravitational wave events will provide unique insights
into the final outcome of core-collapse supernovae and the formation
of black holes.

Comparing our maximum black hole masses to those presented in
\citet{Belczynski16} we find the difference between predicted
black-hole masses is roughly a factor between 1.25 to 2 at
metallicities around Solar and below. The assumed method of
calculating remnant masses from the final stellar model likely
contributes to this, as does our use of detailed evolution models
rather than the rapid models. While in general rapid models are
sufficiently accurate to estimate the evolution of a star, in the
cases of significant mass loss the results can differ to those from
detailed models. For example in cases where the hydrogen envelope of a
massive star is close to being removed a detailed model will end its
evolution as a yellow supergiant, while a rapid model typically
predicts the star becomes a Wolf-Rayet star. In such cases the
eventual mass of the star and compact remnant will be different. In
addition, as discussed above, binary interactions can lead to a star
forming a more massive black hole than it would in single star
evolution. These are only second-order effects and can be accounted
for in rapid models to some degree but can only be revealed by a
detailed evolution model due to the highly non-linear nature of
stellar evolution.

We note that \citet{2016A&A...588A..50M} also used a grid of detailed
evolution models to investigate the evolution of binary stars in tight
orbits that experience QHE driven by tides rather than mass
transfer. They also find a metallicity cutoff similar to that of
\citet{Belczynski16}. However they did not investigate the standard
binary evolution channel that is presented here and that metallicity
limit is for the massive overcontact binary pathway when the two stars
are tidally lock and both experience QHE. We do not include this
pathway in our study.

For a very low metallicity ($Z\le 10^{-4}$), 1/200th Solar, the merger
rates predicted by BPASS are at their highest, a factor of 100 greater
than at higher metallicities. For a stellar population undergoing
constant star formation, the rate of binary black hole merger events
in our models peaks at close to the mass of the black holes inferred
for GW\,150914.  However at more typical stellar metallicities, this
event would constitute a less common high-mass outlier in the
predicted distribution.  The predictions of BPASS are consistent with
this event arising from a normal, if low metallicity, stellar
population. While other scenarios, such as that suggested by
\citet{MandeldeMink2016,2016A&A...588A..50M}, could also lead to the
formation of such a binary, exotic scenarios are not necessarily
required. We note the importance of considering metallicity biases in
the host stellar population when attempting to localise
electromagnetic counterparts for binary merger events.

Finally, while this paper has been under review the LIGO consortium
have announced the detection of a further binary black-hole merger,
and undertaken further analysis of a third, low significance event
\citep{2016arXiv160604856T}. Neither event had a detected optical
counterpart. We have included these events (GW\,151226 and
LVT\,151012) on Figure \ref{figmasses}. Their total black-hole binary
masses of 37 and 22M$_{\odot}$ are lower than that of GW150914 and, as
we can see in Figure \ref{figrates}, rates of similar events are
expected to be higher and more typical of what is expected from binary
black-hole mergers in the Universe. We note with interest that one or
both of the two black holes in GW\,151226 may have had intrinsic
spin. As the population of gravitational wave events grows, this
signal will be key to identifying cases where QHE is important in
producing the progenitor system. Together the three events have also
allowed the LIGO consortium to report a more accurate estimate for the
expected merger rate of 9 to 240~Gpc$^{-3}$~yr$^{-1}$
\citep{2016arXiv160604856T}, comparable to the estimate presented in
this work of 10 to 100~Gpc$^{-3}$~yr$^{-1}$.

\vspace{-10pt}\section*{Acknowledgements} JJE acknowledges support
from the University of Auckland. ERS acknowledges support from UK STFC
consolidated grant ST/L000733/1. The authors thank Richard Easther for
useful discussion. We acknowledge and thank the LIGO Scientific
Collaboration for their decades of endeavours that led to the
detection of GW150914. The authors wish to acknowledge the
contribution of the NeSI high-performance computing facilities and the
staff at the Centre for eResearch at the University of Auckland. New
Zealand's national facilities are provided by the New Zealand eScience
Infrastructure (NeSI) and funded jointly by NeSI's collaborator
institutions and through the Ministry of Business, Innovation and
Employment Infrastructure programme (http://www.nesi.org.nz)



\bsp	
\label{lastpage}

\begin{thebibliography}{99}



\bibitem[\protect\citeauthoryear{Abadie et al.}{2010}]{2010CQGra..27q3001A}
Abadie J., et al., 2010, CQGra, 27, 3001


\bibitem[\protect\citeauthoryear{Abbott et al}{2016a}]{discovery}
Abbott B. P. et al, 2016a, Phys. Rev. Lett. 116, 061102

\bibitem[\protect\citeauthoryear{Abbott et al}{2016b}]{properties}
Abbott B. P. et al, 2016b, Phys. Rev. Lett. submitted, gr-qc: 1602.03840

\bibitem[\protect\citeauthoryear{Abbott et al}{2016c}]{implications}
Abbott B. P. et al, 2016c, ApJ, 818, 2

\bibitem[\protect\citeauthoryear{Abbott et al}{2016d}]{2016arXiv160208492A}
Abbott B. P. et al, 2016d, arXiv:1602.08492

\bibitem[\protect\citeauthoryear{Abbott et al}{2016e}]{2016arXiv160203842A}
Abbott B. P. et al, 2016e, arXiv:1602.03842

\bibitem[\protect\citeauthoryear{Abbott et al}{2016f}]{2016arXiv160604856T}
Abbott B. P. et al, 2016f, arXiv:1606.04856

\bibitem[\protect\citeauthoryear{Allende Prieto, Lambert, 
\& Asplund}{2002}]{2002ApJ...573L.137A} Allende Prieto C., Lambert D.~L., Asplund M., 2002, ApJ, 573, L137 

\bibitem[\protect\citeauthoryear{Asplund}{2005}]{2005ARAA..43..481A} Asplund M., 2005, ARA\&A, 43, 481 

\bibitem[\protect\citeauthoryear{Belczynski \& Taam}{2008}]{2008ApJ...685..400B}
Belczynski K., Taam R.E., 2008, ApJ, 685, 400B


\bibitem[\protect\citeauthoryear{Belczynski et al.}{2010}]{Belczynski10}
Belczynski K., Dominik M., Bulik T., O'Shaughnessy R., Fryer C., Holz D.E., 2010, ApJ, 715, L138

\bibitem[\protect\citeauthoryear{Belczynski et al.}{2012}]{2012ApJ...757...91B}
Belczynski K., Wiktorowicz G., Fryer C.L., Holz D.E., Kalogera V., 2012, ApJ, 757, 91B

\bibitem[\protect\citeauthoryear{Belczynski et al.}{2015}]{Belczynski15}
Belczynski K., Repetto S., Holz,D., O'Shaughnessy R., Bulik T., Berti E., Fryer C., Dominik M., 2016, ApJ, 819, 108

\bibitem[\protect\citeauthoryear{Belczynski et al.}{2016}]{Belczynski16}
  Belczynski K., Holz D.E., Bulik T., O'Shaughnessy R., 2016, arXiv:160204531B

\bibitem[\protect\citeauthoryear{Berry et al.}{2015}]{Berry15}
Berry C.P.L. et al., 2015, ApJ, 804, 114

\bibitem[\protect\citeauthoryear{Bloemen et 
al.}{2015}]{2015ASPC..496..254B} Bloemen S., Groot P., Nelemans G., 
Klein-Wolt M., 2015, ASPC, 496, 254 

\bibitem[\protect\citeauthoryear{Bobrick, Davies \&
    Church}{2015}]{BDC15} Bobrick A., Davies M.B., Church
  R.P., 2015, `Proceedings of the MG13 Meeting on General
  Relativity'. Edited by Rosquist K. et al., Published by World
  Scientific Publishing Co. Pte. Ltd., pp. 1820-1821
	
\bibitem[\protect\citeauthoryear{Bogomazov, Lipunov \& Tutukov}{2007}]{2007ARep...51..308B}
Bogomazov A. I., Lipunov V. M., Tutukov A. V., 2007, ARep, 51, 308B

\bibitem[\protect\citeauthoryear{Bruzual \& Charlot}{2003}]{2003MNRAS.344.1000B} Bruzual G., Charlot S., 2003, MNRAS, 344, 1000 

\bibitem[\protect\citeauthoryear{Carr}{2003}]{2003LNP...631..301C}
Carr B.J., 2003, LNP, 631, 301

\bibitem[\protect\citeauthoryear{Connaughton et al.}{2016}]{2016arXiv160203920C}
Connaughton V. et al., 2016, arXiv: 1602.03920

\bibitem[\protect\citeauthoryear{Crowther et al.}{2010}]{2010MNRAS.403L..41C}
Crowther P. A., Barnard R., Carpano S., Clark J. S., Dhillon V. S., Pollock, A. M. T., 2010, MNRAS, 403L, 41C

\bibitem[\protect\citeauthoryear{Davies et al.,}{2009}]{2009ApJ...707..844D}
Davies B., Figer D. F., Kudritzki R.-P., Trombley C., Kouveliotou C., Wachter S., 2009, ApJ, 707, 844D


\bibitem[\protect\citeauthoryear{de Mink \& Belczynski}{2015}]{deMinkBelczynski2015}
de Mink, S. E.; Belczynski, K.2015ApJ...814...58D

\bibitem[\protect\citeauthoryear{Dominik et al.}{2012}]{2012ApJ...759...52D}	
Dominik M., Belczynski K., Fryer C., Holz D.E., Berti E., Bulik T., Mandel I., O'Shaughnessy, R., 2012, ApJ, 759, 52D

\bibitem[\protect\citeauthoryear{Dominik et al.}{2013}]{Dominik2013}
Dominik M., Belczynski K., Fryer C., Holz D.E., Berti E., Bulik T., Mandel I., O'Shaughnessy R., 2013, ApJ, 779, 72D

\bibitem[\protect\citeauthoryear{Dominik et al.}{2015}]{Dominik2015}
Dominik M., et al, 2015, ApJ, 806, 263D

\bibitem[\protect\citeauthoryear{Dosopoulou \& Kalogera}{2016a}]{DK16a}
Dosopoulou F., Kalogera V., 2016a, ApJ in press

\bibitem[\protect\citeauthoryear{Dosopoulou \& Kalogera}{2016b}]{DK16b}
Dosopoulou F., Kalogera V., 2016b, ApJ in press



\bibitem[\protect\citeauthoryear{Eldridge \& Tout}{2004}]{ET2004}
Eldridge J. J., Tout C. A., 2004, MNRAS, 353, 87E

\bibitem[\protect\citeauthoryear{Eldridge, Izzard \& Tout}{2008}]{EIT2008}
Eldridge J. J., Izzard R. G., Tout C. A., 2008, MNRAS, 384, 1109E

\bibitem[\protect\citeauthoryear{Eldridge}{2009}]{E2009}
Eldridge J. J.,	2009, MNRAS, 400, 20E

\bibitem[\protect\citeauthoryear{Eldridge \& Stanway}{2009}]{ES2009}
Eldridge J. J., Stanway E. R.,	2009, MNRAS, 400, 1019E

\bibitem[\protect\citeauthoryear{Eldridge et al.}{2013}]{2013MNRAS.436..774E}
Eldridge J.J., Fraser M., Smartt S.J., Maund J.R., Crockett R.M.,2013, MNRAS, 436, 774

\bibitem[\protect\citeauthoryear{Eldridge et al.}{2015}]{2015MNRAS.446.2689E}
Eldridge J. J., Fraser M., Maund J. R., Smartt S. J., 2015, MNRAS, 446, 2689E

\bibitem[\protect\citeauthoryear{Eldridge, Langer \& Tout}{2011}]{ELT2011}
Eldridge, J. J., Langer N., Tout C. A., 2011, MNRAS, 414, 3501E

\bibitem[\protect\citeauthoryear{Eldridge \& Stanway}{2012}]{ES2012}
  Eldridge, J. J., Stanway E. R., 2012, MNRAS, 419, 479E

\bibitem[\protect\citeauthoryear{Essick et al.}{2015}]{2015ApJ...800...81E} 
Essick R., Vitale S., Katsavounidis E., Vedovato G., Klimenko S., 2015, 
ApJ, 800, 81 

\bibitem[\protect\citeauthoryear{Gehrels et 
al.}{2015}]{2015arXiv150803608G} Gehrels N., Cannizzo J.~K., Kanner J., 
Kasliwal M.~M., Nissanke S., Singer L.~P., 2015, arXiv, arXiv:1508.03608 

\bibitem[\protect\citeauthoryear{Ghosh et al.}{2015}]{2015arXiv151102673G} 
Ghosh S., Bloemen S., Nelemans G., Groot P.~J., Price L.~R., 2015, arXiv, 
arXiv:1511.02673 

\bibitem[\protect\citeauthoryear{Greiner et al.}{2016}]{2016arXiv160600314G}
Greiner J., Burgess J.M., Savchenko V., Yu H.-F., 2016, ApJL in press.


\bibitem[\protect\citeauthoryear{Heger \& Woosley}{2002}]{2002ApJ...567..532H}
Heger A., Woosley S. E., 2002, ApJ, 567, 532

\bibitem[\protect\citeauthoryear{Heger et al.}{2003}]{2003ApJ...591..288H}
Heger A., Fryer C. L., Woosley S. E., Langer N., Hartmann D. H., 2003, ApJ, 591, 288H

\bibitem[\protect\citeauthoryear{Hobbs et al.}{2005}]{2005MNRAS.360..974H} 
Hobbs G., Lorimer D.~R., Lyne A.~G., Kramer M., 2005, MNRAS, 360, 974 

\bibitem[\protect\citeauthoryear{Hulse 
\& Taylor}{1975}]{1975ApJ...195L..51H} Hulse R.~A., Taylor J.~H., 1975, ApJ, 195, L51 

\bibitem[\protect\citeauthoryear{Hurley, Tout \& Pols}{2002}]{2002MNRAS.329..897H}
Hurley J.R., Tout C.A., Pols O.R., 2002, MNRAS, 329, 897H

\bibitem[\protect\citeauthoryear{James et al.}{2013}]{2013MNRAS.430.2097J} James B.~L., Tsamis Y.~G., Walsh J.~R., Barlow M.~J., Westmoquette M.~S., 2013, MNRAS, 430, 2097 

\bibitem[\protect\citeauthoryear{Kowalska-Leszczynska et al.}{2015}]{Kowalska2015}
Kowalska-Leszczynska I., Regimbau T., Bulik T., Dominik M., Belczynski K., 2015, A\&A, 574A, 58K

\bibitem[\protect\citeauthoryear{Kroupa, Tout, 
\& Gilmore}{1993}]{1993MNRAS.262..545K} Kroupa P., Tout C.~A., Gilmore G., 1993, MNRAS, 262, 545 

\bibitem[\protect\citeauthoryear{Kiminki \& Kobulnicky}{2012}]{2012ApJ...751....4K}
Kiminki D.C., Kobulnicky H.A., 2012, ApJ, 751, 4K

\bibitem[\protect\citeauthoryear{Kulkarni et al.}{2013}]{2013ApJ...772...93K}
Kulkarni G., Rollinde E., Hennawi J. F., Vangioni E., 2013, ApJ, 772, 93

\bibitem[\protect\citeauthoryear{Izzard, Ramirez-Ruiz \& Tout}{2004}]{IRT2004}
Izzard R. G., Ramirez-Ruiz E., Tout C. A., 2004, MNRAS, 348, 1215I

\bibitem[\protect\citeauthoryear{Ivanova et al.}{2013}]{IV13}
Ivanova, N. et al., 2013, A\&ARv, 21, 59I

\bibitem[\protect\citeauthoryear{Leitherer et al.}{1999}]{1999ApJS..123....3L} Leitherer C., et al., 1999, ApJS, 123, 3 

\bibitem[\protect\citeauthoryear{LIGO Scientific Collaboration et 
al.}{2015}]{2015CQGra..32g4001L} LIGO Scientific Collaboration, et al., 
2015, CQGra, 32, 074001 

\bibitem[\protect\citeauthoryear{Lipunov \& Pruzhinskaya}{2014}]{2014MNRAS.440.1193L}
Lipunov V. M., Pruzhinskaya M. V., 2014, MNRAS, 440, 1193L

\bibitem[\protect\citeauthoryear{Ma et al.}{2016}]{2016arXiv160107559M} Ma 
X., Hopkins P.~F., Kasen D., Quataert E., Faucher-Giguere C.-A., Keres D., 
Murray N., 2016, arXiv, arXiv:1601.07559 

\bibitem[\protect\citeauthoryear{Madau \& Dickinson}{2014}]{2014ARA&A..52..415M}	
Madau P., Dickinson M.,	2014, ARA\&A, 52, 415M

\bibitem[\protect\citeauthoryear{Mandel}{2016}]{Mandel2016}	
Mandel I., 2016, MNRAS, 456, 578M

\bibitem[\protect\citeauthoryear{Mandel \& de Mink}{2016}]{MandeldeMink2016}	
Mandel I., de Mink S. E., 2016, arXiv:1601.00007M

\bibitem[\protect\citeauthoryear{Mennekens \& Vanbeveren}{2016}]{Mennekens16}
Mennekens N., Vanbeveren, D., 2016, arXiv:160106966M

\bibitem[\protect\citeauthoryear{Marchant et al.}{2016}]{2016A&A...588A..50M}
Marchant P., Langer N., Podsiadlowski P., Tauris T. M., Moriya T. J., 2016, A\&A, 588A, 50M

\bibitem[\protect\citeauthoryear{Nieva \& Przybilla}{2012}]{2012AA...539A.143N}
Nieva M.-F., Przybilla N., 2012, A\&A, 539A, 143N

\bibitem[\protect\citeauthoryear{\"{O}zel et al.}{2010}]{2010ApJ...725.1918O}
\"{O}zel F., Psaltis D., Narayan R., McClintock J.E., 2010, ApJ, 725, 1918O

\bibitem[\protect\citeauthoryear{Peters}{1964}]{1964PhRv..136.1224P} Peters 
P.~C., 1964, PhRv, 136, 1224 

\bibitem[\protect\citeauthoryear{Podsiadlowski}{1992}]{1992PASP..104..717P}
Podsiadlowski P., 1992, PASP, 104, 717P

\bibitem[\protect\citeauthoryear{Sana et al.}{2012}]{Sana2012}
Sana H. et al., 2012, Science, 337, 444

\bibitem[\protect\citeauthoryear{Sana et al.}{2014}]{Sana2014}
Sana H. et al., 2014, ApJS, 215, 15S

\bibitem[\protect\citeauthoryear{Singer et al.}{2014}]{2014ApJ...795..105S} 
Singer L.~P., et al., 2014, ApJ, 795, 105 

\bibitem[\protect\citeauthoryear{Spera, Mapelli \& Bressan}{2015}]{2015MNRAS.451.4086S}
Spera M., Mapelli M., Bressan A., 2015, MNRAS, 451, 4086S

\bibitem[\protect\citeauthoryear{Stanway et al.}{2014}]{2014MNRAS.444.3466S} Stanway E.~R., Eldridge J.~J., Greis S.~M.~L., Davies L.~J.~M., Wilkins S.~M., Bremer M.~N., 2014, MNRAS, 444, 3466 

\bibitem[\protect\citeauthoryear{Stanway, Eldridge \& Becker}{2016}]{SEB2016}
  Stanway E. R., Eldridge J. J., Becker G. D., 2016, MNRAS, 456, 485S

\bibitem[\protect\citeauthoryear{Sukhbold et al.}{2016}]{2015arXiv151004643S}
Sukhbold T., Ertl T., Woosley S.E.., Brown J.M., Janka H.-T., 2016, ApJ in press.

\bibitem[\protect\citeauthoryear{Tinsley 
\& Gunn}{1976}]{1976ApJ...203...52T} Tinsley B.~M., Gunn J.~E., 1976, ApJ, 203, 52 

\bibitem[\protect\citeauthoryear{Tremonti et 
al.}{2004}]{2004ApJ...613..898T} Tremonti C.~A., et al., 2004, ApJ, 613, 
898 

\bibitem[\protect\citeauthoryear{Turner et al.}{2014}]{2014MNRAS.445..794T} Turner M.~L., Schaye J., Steidel C.~C., Rudie G.~C., Strom A.~L., 2014, MNRAS, 445, 794 

\bibitem[\protect\citeauthoryear{Tutukov \& Yungelson}{1973}]{TY73}
Tutukov A., Yungelson L., 1973, Nauchnye Informatsii, 27, 70

\bibitem[\protect\citeauthoryear{Ugliano et al.}{2012}]{2012ApJ...757...69U}
Ugliano M., Janka H.-T., Marek A., Arcones A., 2012, ApJ, 757, 69

\bibitem[\protect\citeauthoryear{Vanbeveren, De Loore \& Van Rensbergen}{1998}]{vanbev1998} Vanbeveren D., De Loore C., Van Rensbergen W., 1998, A\&ARv, 9, 63V

\bibitem[\protect\citeauthoryear{Villante et al.}{2014}]{2014ApJ...787...13V}
Villante F. L., Serenelli A.M., Delahaye F., Pinsonneault M.H., 2014, ApJ, 787, 13V

\bibitem[\protect\citeauthoryear{Vink, de Koter \& Lamers}{2001}]{Vinketal}
Vink J. S., de Koter A., Lamers H. J. G. L. M., 2001, A\&A, 369, 574V

\bibitem[\protect\citeauthoryear{Walborn et al.}{1987}]{1987ApJ...321L..41W}
Walborn N. R., Lasker B. M., Laidler V. G., Chu Y.-H., 1987, ApJ, 321L, 41

\bibitem[\protect\citeauthoryear{White, Daw, 
    \& Dhillon}{2011}]{2011CQGra..28h5016W} White D.~J., Daw E.~J., Dhillon V.~S., 2011, CQGra, 28, 085016

\bibitem[\protect\citeauthoryear{Wilkins et 
al.}{2015}]{2015arXiv151203214W} Wilkins S.~M., Feng Y., Di Matteo T., 
Croft R., Stanway E.~R., Bouwens R.~J., Thomas P., 2015, arXiv, 
arXiv:1512.03214 


\bibitem[\protect\citeauthoryear{Wofford et 
al.}{2016}]{2016arXiv160103850W} Wofford A., et al., 2016, arXiv, 
arXiv:1601.03850 
	
\bibitem[\protect\citeauthoryear{Wiktorowicz et al.}{2015}]{2015ApJ...810...20W}
Wiktorowicz G., Sobolewska M., Sa˛dowski A., Belczynski K., 2015, ApJ, 810, 20W

\bibitem[\protect\citeauthoryear{Wyrzykowski et al.}{2016}]{microlense}
Wyrzykowski L. et al., 2016, MNRAS, 458, 3012


\bibitem[\protect\citeauthoryear{Xiao \& Eldridge}{2015}]{2015MNRAS.452.2597X}
Xiao L., Eldridge J.J., 2015, MNRAS, 452, 2597

\bibitem[\protect\citeauthoryear{Yoon, Langer \& Norman}{2006}]{2006A&A...460..199Y}
Yoon S.-C., Langer N., Norman C., 2006, A\&A, 460, 199Y

\bibitem[\protect\citeauthoryear{Yusof et al.}{2013}]{2013MNRAS.433.1114Y}
Yusof N., Hirschi R., Meynet G., Crowther P.A., Ekström S., Frischknecht U., Georgy C., Abu Kassim H., Schnurr O., 2013, MNRAS, 433, 1114Y







\end{thebibliography}
\end{document}